\let\includefigures=\iftrue

\input harvmac
\noblackbox
\includefigures
\message{If you do not have epsf.tex (to include figures),}
\message{change the option at the top of the tex file.}
\input epsf
\def\figin{\epsfcheck\figin}\def\figins{\epsfcheck\figins}
\def\epsfcheck{\ifx\epsfbox\UnDeFiNeD
\message{(NO epsf.tex, FIGURES WILL BE IGNORED)}
\gdef\figin##1{\vskip2in}\gdef\figins##1{\hskip.5in}
\else\message{(FIGURES WILL BE INCLUDED)}%
\gdef\figin##1{##1}\gdef\figins##1{##1}\fi}
\def\DefWarn#1{}
\def\figinsert{\goodbreak\midinsert}
\def\ifig#1#2#3{\DefWarn#1\xdef#1{fig.~\the\figno}
\writedef{#1\leftbracket fig.\noexpand~\the\figno}%
\figinsert\figin{\centerline{#3}}\medskip\centerline{\vbox{
\baselineskip12pt\advance\hsize by -1truein
\noindent\footnotefont{\bf Fig.~\the\figno:} #2}}
\bigskip\endinsert\global\advance\figno by1}
\else
\def\ifig#1#2#3{\xdef#1{fig.~\the\figno}
\writedef{#1\leftbracket fig.\noexpand~\the\figno}%
\global\advance\figno by1} \fi

\lref\ggl{H. Georgi and S.L. Glashow, Phys. Rev. Lett. {\bf 32} (1974) 438.}%
\lref\dg{S. Dimopoulos and H. Georgi, Nucl. Phys. {\bf B192}
(1981) 150.}%
\lref\RSAdS{J. Maldacena, unpublished; E. Witten, unpublished;  S.
Gubser, hep-th/9912001;
M. Duff and J. Liu, Phys. Rev. Lett. {\bf 85} (2000) 2052,
hep-th/0003237; N. Arkani-Hamed, M. Porrati and L. Randall,
hep-th/0012148; R. Rattazzi and A. Zaffaroni, hep-th/0012248. }

\lref\hs{M. Dine, W. Fischler, Nucl. Phys. {\bf B204} (1982) 346;
S. Dimopoulos and S. Raby, Nucl. Phys. {\bf B219} (1982) 479; J.
Polchinski and L. Susskind, Phys. Rev. {\bf D26} (1982) 3661.}
\lref\drw{S. Dimopoulos, S. Raby and F. Wilczek, Phys. Rev. {\bf
D24} (1981) 1681.} \lref\noscale{J. Ellis, C. Kounnas and D.
Nanopoulos, Nucl. Phys. {\bf B247} (1984) 373.} \lref\pomarol{A.
Pomarol, Phys. Lett. {\bf B486} (2000) 153; Phys. Rev. Lett. {\bf
85} (2000) 4004; T. Gherghetta and A. Pomarol, Nucl. Phys. {\bf
B586} (2000) 141, hep-ph/0012378.} \lref\suGUT{ E. Cremmer, S.
Ferrara, L. Girardello, and A. Van Proyen, Phys. Lett. {\bf B116}
(1982) 231; A. Chamseddine, R. Arnowitt, and P. Nath, Phys. Rev.
Lett. {\bf 49} (1982) 970; R. Barbieri, S. Ferrara, and C. Savoy,
Phys. Lett. {\bf B110} (1982) 343; L. J. Hall, J. Lykken, and S.
Weinberg, Phys. Rev. {\bf D27}(1983) 2359.}

\lref\rszero{L. Randall  and R. Sundrum, Nucl. Phys. {\bf B557}
(1999) 79; G.F. Giudice, M.A. Luty, H. Murayama and R. Rattazzi,
JHEP {\bf 9812} (1998) 027.}
\lref\mart{D.~E.~Kaplan, G.~D.~Kribs and M.~Schmaltz,
Phys.\ Rev.\ {\bf D62}, 035010 (2000);
Z.~Chacko, M.~A.~Luty, A.~E.~Nelson and E.~Ponton,
JHEP {\bf 0001}, 003 (2000);
M.~Schmaltz and W.~Skiba,
Phys.\ Rev.\ {\bf D62}, 095005 (2000);
Phys.\ Rev.\  {\bf D62}, 095004 (2000);
D.~E.~Kaplan and G.~D.~Kribs,
JHEP {\bf 0009}, 048 (2000).}
\lref\nil{H.P. Nilles, Phys. Lett. {\bf B115} (1982) 193; Nucl.
Phys. {\bf B217} (1983) 366.} \lref\kaplu{V.S. Kaplunovsky and J.
Louis, Phys. Lett. {\bf B306} (1993) 269.}

\lref\gauco{M. Dine, R. Rohm, N. Seiberg, E. Witten, Phys. Lett.
{\bf B156} (1985) 55.}

\lref\juan{J. Maldacena, ATMP {\bf 2} (1998) 231, hep-th/9711200;
S. Gubser, I. Klebanov and A. Polyakov, Phys. Lett. {\bf B428} (1998)
105, hep-th/9802109; E. Witten, ATMP {\bf 2} (1998) 253, hep-th/9802150.}

\lref\RSI{L. Randall and R. Sundrum, Phys. Rev. Lett. {\bf 83}
(1999) 3370, hep-th/9905221.} \lref\AdSCFT{S. Gubser, I. Klebanov
and A. Polyakov, Phys. Lett. {\bf B428} (1998) 105,
hep-th/9802109; E. Witten, ATMP {\bf 2} (1998) 253,
hep-th/9802150.} \lref\albion{V. Balasubramanian, P. Kraus and A.
Lawrence, Phys. Rev. {\bf D59} (1999) 046003, hep-th/9805171.}
\lref\dimo{E. Kolb and M. Turner, {\it The Early Universe},
Addison-Wesley, Redwood City, 1990.} \lref\spesta{ D.N. Spergel
and P.J. Steinhardt, Phys. Rev. Lett. {\bf 84} (2000) 3760,
astro-ph/9909386.} \lref\peebles{P.J.E. Peebles,
astro-ph/0002495.} \lref\dine{M. Dine and A. Nelson, Phys. Rev.
{\bf D48} (1993) 1277, hep-th/9303230.}

\lref\cen{R. Cen, astro-ph/0005206.}
\lref\ygmn{Y. Grossman and M.
Neubert, Phys. Lett. {\bf B474} (2000) 361, hep-ph/9912408}
\lref\joe{ J.~Dai, R.~G.~Leigh and J.~Polchinski,
Mod.\ Phys.\ Lett.\  {\bf A4}, 2073 (1989); P.~Horava,
Phys.\ Lett.\  {\bf B231}, 251 (1989); P.~Horava,
Nucl.\ Phys.\  {\bf B327}, 461 (1989); G.~Pradisi and A.~Sagnotti,
Phys.\ Lett.\  {\bf B216}, 59 (1989); J.~Polchinski,
Phys.\ Rev.\ Lett.\ {\bf 75}, 4724 (1995), hep-th/9510017. }
\lref\dfgk{O. DeWolfe, D. Freedman, S. Gubser and A. Karch,
``Modeling the Fifth-Dimension with Scalars and Gravity,'' Phys.
Rev. {\bf D62} (2000) 046008.}
\lref\add{N. Arkani-Hamed, S.
Dimopoulos and G. Dvali, Phys. Lett. {\bf B429} (1998) 263,
hep-th/9803315; I. Antoniadis, N. Arkani-Hamed, S. Dimopoulos and
G. Dvali, Phys. Lett. {\bf B436} (1998) 257, hep-th/9804398; N.
Arkani-Hamed, S. Dimopoulos and G. Dvali, Phys. Rev. {\bf D59}
(1999) 086004, hep-th/9807344.}
\lref\dkklsII{S. Dimopoulos, S.
Kachru, N. Kaloper, A. Lawrence and E. Silverstein,
hep-th/0106128. }
\lref\shiu{G. Shiu and S.H. Tye, ``TeV Scale Superstring
and Extra Dimensions,'' Phys. Rev. {\bf D58} (1998) 106007,
hep-th/9805157.}
\lref\hw{P. Ho\v rava and E. Witten, Nucl. Phys.
{\bf B460} (1996) 506, hep-th/9510209.}
\lref\HV{H. Verlinde,
Nucl. Phys. {\bf B580} (2000) 264, hep-th/9906182; S. Giddings,
S. Kachru and J. Polchinski, preprint to appear.}
\lref\edunify{E. Witten, Nucl. Phys. {\bf B471} (1996) 135,
hep-th/9602070.}
\lref\svw{S. Sethi, C. Vafa and E. Witten,
``Constraints on Low Dimensional String Compactifications,'' Nucl.
Phys. {\bf B480} (1996) 213, hep-th/9606122.}

\lref\sutter{S. Dimopoulos and D. Sutter, Nucl. Phys. {\bf B452} (1995) 496.}

\lref\ovrut{For a review with references, see: A. Lukas, B. Ovrut
and D. Waldram, ``Heterotic M-theory Vacua with Five-branes,''
Fortsch. Phys. {\bf 48} (2000) 167, hep-th/9903144.}

\lref\braneworld{Recent attempts to construct such models can be
found in: G. Aldazabal, S. Franco, L.E. Ibanez, R. Rabadan and
A.M. Uranga, ``D=4 Chiral String Compactifications from
Intersecting Branes,'' hep-th/0011073; G. Aldazabal, S. Franco,
L.E. Ibanez, R. Rabadan and A.M. Uranga, ``Intersecting Brane
Worlds,'' hep-ph/0011132; G. Aldazabal, L.E. Ibanez, F. Quevedo
and A.M. Uranga, ``D-branes at Singularities: A Bottom Up Approach
to the String Embedding of the Standard Model,'' JHEP {\bf 0008}
(2000) 002, hep-th/0005067.}

\lref\adsbreak{S. Kachru and E. Silverstein, ``4d Conformal Field
Theories and Strings on Orbifolds,'' Phys. Rev. Lett. {\bf 80}
(1998) 4855, hep-th/9802183\semi A. Lawrence, N. Nekrasov and C.
Vafa, ``On Conformal Field Theories in Four Dimensions,'' Nucl.
Phys. {\bf B533} (1998) 199, hep-th/9803015\semi J. Distler and F.
Zamora, ``Non-Supersymmetric Conformal Field Theories from Stable
Anti-de Sitter Spaces,'' Adv. Theor. Math. Ph. {\bf 2} (1999)
1405, hep-th/9810206.}

\lref\bklt{V. Balasubramanian, P. Kraus, A. Lawrence and S.
Trivedi, ``Holographic probes of anti-de Sitter spacetimes,''
Phys. Rev. {\bf D59}\ (1999) 104021; hep-th/9808017.} \lref\bgl{V.
Balasubramanian, S.B. Giddings and A. Lawrence, ``What do CFTs
tell us about anti-de Sitter spacetimes?'' JHEP {\bf 9903}\ (1999)
001; hep-th/9902052.} \lref\bdhm{T. Banks, M.R. Douglas, G.T.
Horowitz and E. Martinec, ``AdS dynamics from conformal field
theory,'' hep-th/9808016.} \lref\scatt{I.R. Klebanov, ``World
volume approach to absorption by non-dilatonic branes,'' Nuc.
Phys. {\bf B496}\ (1997) 231.} \lref\brfreed{P. Breitenlohner and
D.Z. Freedman, ``Positive energy in anti-de Sitter backgrounds and
gauged extended supergravity,'' Phys. Lett. {\bf B115}\ (1982)
197; ibid., ``Stability in gauged extended supergravity,'' Ann.
Phys. NY {\bf 144}\ (1982) 249; L. Mezinescu and P.K. Townsend,
``Stability at a local maximum in higher-dimensional anti-de
Sitter space and applications to supergravity,'' Ann. Phys. NY
{\bf 160}\ (1985) 406.} \lref\Igor{S. Gubser, I. Klebanov, and A.
Peet, ``Entropy and Temperature of Black 3-Branes'', Phys. Rev.
{\bf D54} (1996) 3915, hep-th/9602135.} \lref\IgorII{I. Klebanov,
``Wordvolume Approach to Absorption by Nondilatonic Branes'',
Nucl. Phys. {\bf B496} (1997) 231; S. Gubser, I. Klebanov, A.
Tseytlin, ``String Theory and Classical Absorption by
Three-Branes'', Nucl. Phys. {\bf B499} (1997) 217.}
\lref\gravbox{J. Lykken, R. Myers, J. Wang, ``Gravity in a Box'',
JHEP {\bf 0009} (2000) 009.} \lref\mirpe{E.A. Mirabelli and M.E.
Peskin, Phys. Rev. {\bf D58} (1998) 065002, hep-th/9712214.}

\lref\appsym{N. Arkani-Hamed and S. Dimopoulos, hep-ph/9811353; N.
Arkani-Hamed, S. Dimopoulos, G. Dvali and J. March-Russell,
hep-ph/9811448; N. Arkani-Hamed and M. Schmaltz, Phys. Rev. {\bf
D61} (2000) 033005, hep-ph/9903417; N. Arkani-Hamed, Y. Grossman
and M. Schmaltz, Phys. Rev. {\bf D61} (2000) 115004,
hep-ph/9909411.}
\lref\selftuning{E. Verlinde and H. Verlinde, JHEP {\bf 0005} (2000) 034;
N. Arkani-Hamed, S. Dimopoulos,
N. Kaloper and R. Sundrum, Phys. Lett. {\bf B480} (2000) 193,
hep-th/0001197; S. Kachru, M. Schulz and E. Silverstein, Phys.
Rev. {\bf D62} (2000) 045021, hep-th/0001206; {it ibid.}, Phys.
Rev. {\bf D62} (2000) 085003;
R. Bousso  and J. Polchinski, JHEP {\bf 0006} (2000) 006;
J.L. Feng, J. March-Russell, S. Sethi and F. Wilczek, hep-th/0005276. }
\lref\heterotic{D.J. Gross, J.A.
Harvey, E. Martinec and R. Rohm, Phys. Rev. Lett {\bf 54} (1985)
502; Nucl. Phys. {\bf B256} (1985) 253; Nucl. Phys. {\bf B267}
(1986) 75.}


\def\frac#1#2{{#1 \over #2}}
\def\p{\partial}

\Title{\vbox{\baselineskip12pt\hbox{hep-th/0104239}
\hbox{SLAC-PUB-8819}\hbox{SU-ITP-00/36}
\hbox{NSF-ITP-01-60}
\hbox{DUKE-CGTP-01-06}}}
{\vbox{ \centerline{Small Numbers From Tunnelling Between}
\bigskip
\centerline{Brane Throats} }}
\bigskip
\centerline{Savas Dimopoulos$^{1}$, Shamit Kachru$^{1,2}$, Nemanja
Kaloper$^{1,2}$,} \centerline{Albion Lawrence$^{1,3}$ and Eva
Silverstein$^{1,2}$}
\bigskip
\centerline{$^{1}${\it Department of Physics and SLAC, Stanford
University, Stanford, CA 94305/94309}} \centerline{$^{2}${\it
Institute for Theoretical Physics, University of California, Santa
Barbara, CA 93106}} \centerline{$^{3}${\it Center for Geometry and
Theoretical Physics, Duke University, Durham, NC 27708}}

\bigskip
\noindent Generic classes of string compactifications include
``brane throats'' emanating from the compact dimensions and
separated by effective potential barriers raised by the background
gravitational fields. The interaction of observers inside
different throats occurs via tunnelling and is consequently weak.
This provides a new mechanism for generating small numbers in
Nature. We apply it to the hierarchy problem, where supersymmetry
breaking near the unification scale causes TeV sparticle masses
inside the standard model throat. We also design naturally
long-lived cold dark matter which decays within a Hubble time to
the approximate conformal matter of a long throat. This may soften
structure formation at galactic scales and raises the possibility
that much of the dark matter of the universe is conformal matter.
Finally, the tunnelling rate shows that the coupling between
throats, mediated by bulk modes, is stronger than a naive
application of holography suggests.

\bigskip
\Date{April 2001}

\newsec{Introduction}

The enormous differences in scales that appear in Nature present a
formidable challenge for any unified theory of forces.  Grand
unification addresses this problem by postulating an energy desert
separating the gravitational and the electroweak scale \ggl. The
supersymmetric version of this picture  \dg, the supersymmetric
standard model (SSM), has had a quantitative success: the
unification prediction of the value of the weak mixing angle \dg,
subsequently confirmed by the LEP and SLC experiments.  While this
picture is attractive, it leaves many fundamental questions
unanswered. There are 125 unexplained parameters, many of them
mysteriously small; these include the masses of the three
generations of particles and the cosmological constant. String
theory provides a natural framework for addressing these
questions. Many scenarios for string phenomenology involve {\it
localized} gauge fields.  Perhaps the simplest is the minimal
Ho\v{r}ava-Witten theory \refs{\hw,\edunify}; other models use
``D-brane'' defects on which gauge dynamics occurs \joe. A
striking possibility emerging from these ingredients is a new
explanation the weakness of gravity \refs{\add}. These ideas are
providing new avenues for exploring physics beyond the Standard
Model, and novel mechanisms for explaining small numbers
\refs{\appsym,\RSI,\selftuning}.

Ho\v{r}ava-Witten theory and the perturbative $E_8 \times
E_8$ heterotic string \heterotic\ have been well studied
in calculable, weakly-coupled
regimes. In this note we will study string phenomenology in a
different calculable regime, which can arise when there are many
branes transverse to the compactification manifold $M$. The
tension of the branes curves the space around them. The
backreaction is proportional to the sum of brane tensions, and
therefore to the total number of branes in some region of space.
Hence solitary branes have little effect and their neighborhood is
nearly flat. Such ``dilute gases'' of branes are commonly studied
in e.g. perturbative string orientifold constructions. In other
regimes of couplings where a (super)gravity description is valid,
large stacks of branes in the compactification manifold $M$
significantly alter the metric on $M$. The regions of space where
the branes reside may be viewed as gravitational funnels, or
throats. Examples in this regime arise in F-theory
compactifications on elliptic Calabi-Yau fourfolds \HV.  From the
4d point of view the geometry is ``warped'' -- the scale factor of
the 4d metric depends on the distance down the throat.
\ifig\octopus{The Calabi-Yau octopus.  $N_i$ is the number of
branes in a given region.} {\epsfxsize3.0in\epsfbox{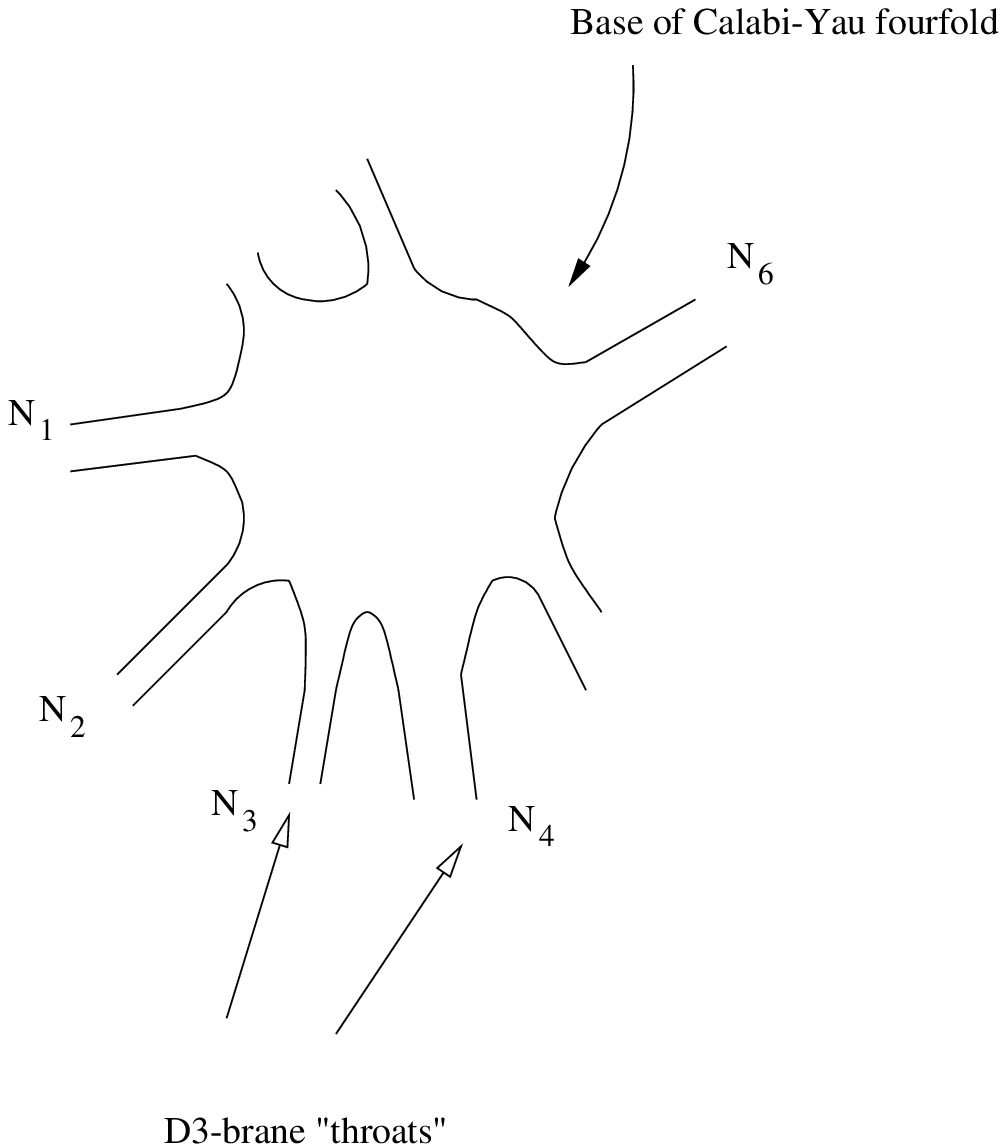}}

In such models, the ensuing geometry of the compactification
resembles an ``octopus,''
where the legs represent throats arising from stacks of branes, as
depicted in Fig. 1. The (super)gravity modes in the throat and the
low-energy field theory on the branes are dual to each other
\juan: the degrees of freedom localized at the ends of the throats
are dual to infrared (IR) excitations of the field theory, while
the excitations closer to the mouth of the throat are dual to the
ultraviolet (UV) degrees of freedom.

This geometry suggests a new mechanism for generating small
numbers in 4d physics. The mutual couplings of the IR degrees of
freedom residing in different throats are suppressed, as these
modes must tunnel through the bulk to communicate. In this paper,
we make this intuition precise in a 5d toy model of Fig. 1, which
appears in Fig. 2. We study a pair of brane throats that are
joined at a ``UV brane'' playing the role of the bulk of $M$. We
then show that the KK modes of the 5d gravity theory localized in
adjacent throats must tunnel to communicate with each other; this
effect can generate small numbers. We study applications to SUSY
breaking and to astrophysical dark matter.
For simplicity and to facilitate a holographic interpretation we will
take AdS metrics in the throats of our toy model.  The effects
we study would persist with much more generic warped metrics (including
those with only power-law warping).
This note is a summary
of \dkklsII, where detailed derivations appear.

\ifig\twoface{A cartoon of two brane throats.}
{\epsfxsize3.0in\epsfbox{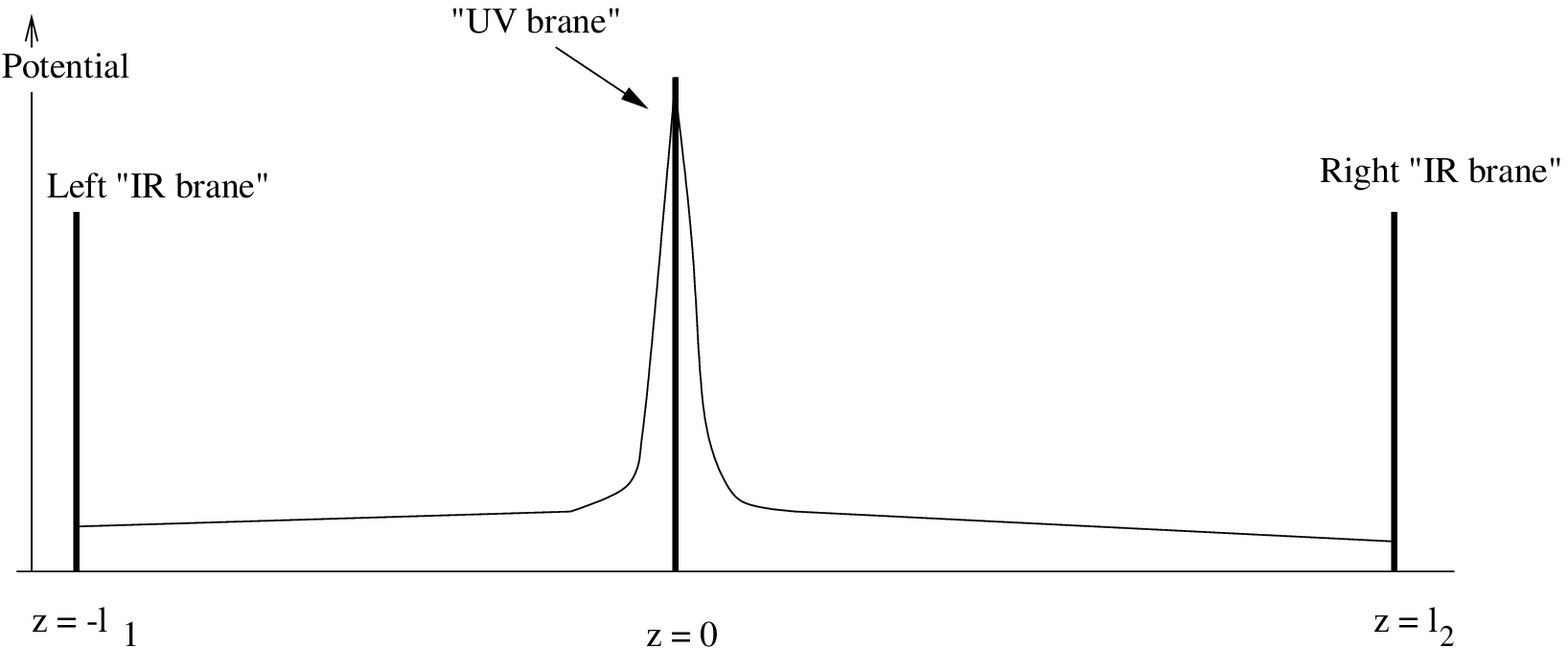}}


\newsec{Tunnelling, Glueball Decay, and Dark Matter}
\medskip
\noindent{\it{The Tunnelling Calculation}}

To get the model depicted in Fig. 2, we choose a 5d metric:
\eqn\metric{ds^2 = {L^2 \over (\vert z \vert + L)^2}
(\eta_{\mu\nu} dx^{\mu}dx^{\nu} + dz^2), ~~-l_1 \leq z \leq l_2\
.} Here $x^\mu$ are the coordinates of our 4d world, and $z$ is
the coordinate down the throat. $l_1 = L e^{R_1/L}$ and $l_2 = L
e^{R_2/L}$, and $R_{1,2}$ are the proper distances from the UV
brane to the left/right IR branes. To analyze the KK spectrum,
define \RSI\ \eqn\newdef{h_{\mu\nu}(x,z) = \sqrt{L \over {\vert z
\vert + L}} e^{i p \cdot x} \psi_{\mu\nu}(z)} where $h_{\mu\nu}$
is the 5d graviton. The transverse, traceless modes of
$h_{\mu\nu}$ satisfy \eqn\kkeq{\p_z^2 \psi + (m^2 - {15 \over {4
(\vert z \vert + L)^2}}) \psi = 0,~~-l_1 \leq z \leq l_2}
with appropriate boundary conditions at the branes,
where
$p^2 = m^2$ is the 4d KK mass of the mode. This is an effective
Schr\"odinger equation with a potential barrier arising from the
warped metric \metric. The 4d and 5d Planck masses scale out this
equation. We expect the low-lying modes in the left/right throat
to have masses $\sim 1/l_{1,2}$, so they must tunnel to
communicate.

Qualitatively similar barriers arise for any minimally coupled
modes in non-AdS backgrounds.  The only differences are the
explicit relationship between the proper bulk distances and the
conformal distances $z$, and between the parameters of the
potential and the bulk scales. Hence our analysis should carry
over to those cases. A precise calculation of the tunneling
amplitude, discussed in \dkklsII, then yields
\eqn\rateag {\Gamma \sim m (mL)^{4}\, .}

\medskip
\noindent{\it{Holographic Interpretation}}

The AdS/CFT correspondence \juan\ states that each AdS throat can
be viewed as a large $N$ gauge theory at strong 't Hooft coupling.
The IR branes provide a schematic representation of confinement or
a mass gap. However, when applying this correspondence to our
background, the usual UV/IR relationship becomes more complicated:
there are light degrees of freedom localized in {\it each} throat,
as well as light ``closed string modes'' localized in the bulk of
$M$. The light KK modes in each throat can be thought of as
``glueballs'' of the dual large $N$ gauge theories. Therefore, eq.
\rateag\ is the rate at which the heavier glueballs of
one strongly coupled gauge theory decay to the lighter glueballs
of the other.

${\it {A~priori}}$, one might have expected our model to be dual
to a system of two gauge theories coupled to each other via 4d
gravity (naively extending the suggestions in \RSAdS). The term in
the interaction Lagrangian inducing glueball decay would be the
dimension 8 operator \eqn\lint{{\cal L}_{int} \sim {1\over
M_{4}^4} T_{\mu\nu}^{(L)} T^{\mu\nu(R)}\ ,} where $T^{(L,R)}$ are
the stress tensors of the left and right dual gauge theories. Such
an interaction would yield a decay rate \eqn\expec{\Gamma \sim
{m^9 \over M_{4}^8}\, \ .}

Instead, Eq. \rateag\ is consistent with a coupling of excitations
in the two throats via a {\it dimension 6} operator, at the scale
$M_{UV} = 1/L$.
Thus the 4d holographic description of our background geometry
consists of two gauge theories coupled by KK modes, at scales set
by the compactification geometry -- in this case, $1/L$.  This
lower scale and the lower dimension of the operator mean that the
inter-throat coupling in these backgrounds can be much larger than
that induced by the coupling to 4d gravity. This is an example of
the more general fact that the effective field theories arising
from string compactifications can have several relevant scales set
by the compactification geometry, beyond just $M_{4}$ which
depends only on the overall volume of $M$.

\medskip
\noindent{\it{CFT Dark Matter}}

It is fascinating to contemplate the possibility that the dark
matter which constitutes about 90 \% of the mass of the universe
is described by a CFT, and that we are immersed inside an ocean of
scale-free matter.\foot{This possibility has been entertained
independently by many physicists, including T. Banks, M. Dine, and
J. Maldacena.} In its simplest form this idea is in conflict with
observation: CFT matter would have a relativistic equation of
state, acting as hot dark matter (HDM); but the large scale
structure of the universe suggests that non-relativistic, cold
dark matter (CDM) dominates the dynamics of the universe since
$t\sim {10^4}$ years. A way to bypass this difficulty is to
postulate that the universe has, until recently, been dominated by
an unstable CDM particle which decays into CFT matter, with a
lifetime of order of the age of the universe. The two-throat model
of Fig. 2 provides such a scenario \dkklsII.  Suppose the SM is
localized on the left IR brane, and that the right throat is dual
to a real CFT (i.e. $l_2 \to \infty$). Now introduce a bulk
particle, the ``bulky,'' which is distinct from the graviton and
which has a bulk parity symmetry under which it changes sign.
This symmetry protects it from decay to SM fields. One can show
\dkklsII\ along the lines of \dimo\ that the relic abundance of
bulkies today would be \eqn\abund{{\rho \over \rho_{c}} \sim
16\pi^2 \times 10^{-2} (M_5 L)^2 {m^2 \over TeV^2} \, .} So if $LM_5 \sim 10$
and $m \sim 100 GeV$, then the left bulkies would close the Universe.

The left bulkies will decay into their much lighter right cousins
with a rate given by Eq. \rateag. A lifetime of the age of the
Universe requires $L^{-1} \sim 10^{14} GeV$ and $M_5 \sim 10^{15}
GeV$. These scales are of the order of the unification scale and
should arise naturally in model building.

In this scenario, the dark matter is slowly decaying into CFT
degrees of freedom in our epoch.  This can have important
observational effects: it can lead to a softening of the dark
matter density profile within our galactic halo, by spreading it
into extragalactic space, as shown in simulations of decaying CDM
performed by Cen \cen. This may help account for the absence of
the excess small scale structure predicted in the canonical CDM
scenario \refs{\cen,\spesta,\peebles}. A variation of this
scenario is to add a right brane with characteristic scale of less
than the galactic halo size, $\sim 400 {\rm kpc}$.  This may
confine most of the approximately conformal matter within the
halo.

\newsec{Tunnelling Mediated Supersymmetry Breaking}

Low-energy SUSY is one of the most attractive scenarios for
physics beyond the Standard Model, because it stabilizes scalar
masses at the SUSY breaking scale \dg. We must still explain
the origin of the low SUSY breaking
scale and the 125 physical parameters in the
MSSM \sutter. There is a variety of SUSY breaking mechanisms
such as gravity \refs{\suGUT,\nil,\gauco}, gauge \refs{\hs,\dine},
anomaly \rszero\ and gaugino \mart\
mediation in hidden sector scenarios. Tunneling
effects between brane throats provide a new mechanism for
generating a small SUSY breaking scale. We will present some basic
results of this approach here, leaving the detailed exposition for
\dkklsII.

Our scenario is as follows. Both IR branes are close to the UV
brane. SUSY is broken on the left IR brane via a soft, R-symmetry
breaking, Majorana-like mass term for bulk fermions. This induces
SUSY breaking mass splittings on the right IR brane, where the SSM
resides. A hierarchy is generated because SUSY breaking at a high
scale on the left IR brane induces small SUSY breaking mass
splittings in the SSM, due to tunnelling supression. We choose the
distance between the right IR brane and the UV brane, as well as
the bulk parameters $M_5$ and $L$, to be near the GUT scale
$M_{GUT} \sim 10^{16} GeV$. This ensures that the cutoff on the
right SSM brane is $M_{GUT}$; consequently supersymmetric gauge
coupling unification \refs{\dg,\drw}\ can be preserved.

We will outline the calculation of the SSM gaugino masses in
\dkklsII.  The action for the bulk fermions is:
\eqn\fermacred{\eqalign{ S_F &= \int d^4x dz \Bigl\{ X^\dagger i
\sigma^\mu \partial_\mu X + Z^\dagger i \bar \sigma^\mu
\partial_\mu Z + {1\over 2} X^\dagger
\partial_z^{\leftrightarrow} Z - {1\over 2} Z^\dagger
\partial_z^{\leftrightarrow} X
    - M_F a(X^\dagger Z + Z^\dagger X)\cr
&~~~~~~~~~~~~~~~~ + \delta(z+l_1) \Bigl( q_1 [ Z^T i\sigma^2 Z -
Z^\dagger i\sigma^2 Z^*] - q_2 [ X^T i\sigma^2 X -  X^\dagger
i\sigma^2 X^*] \Bigr)\Bigr\} \ .}} Here $a^2$ is the warp factor
in \metric. $X$ and $Z$ are bulk spinors, which we have rescaled
by powers of $a$ to have canonical kinetic terms. $M_F$ is a bulk
Dirac mass for the fermion and its superpartner. The SUSY breaking
is encoded in the dimensionless parameters $q_1$ and $q_2$, both
less than unity,
which split the bulk fermions from their superpartners.
Defining: \eqn\matrices{\Sigma= \left(\matrix{~Z~\cr
~X~\cr}\right)\ , } the equations of motion for $\Sigma$ have
solutions of the form: \eqn\bulkfersol{\eqalign{ & \Sigma_L =
\sqrt{w} \left(\matrix{ ~A_L J_{\nu+1/2}(w) + B_L J_{-\nu -
1/2}(w)~ \cr ~A_L J_{\nu-1/2}(w) - B_L J_{-\nu + 1/2}(w)~ \cr}
\right) ~~~~~~~~ z < 0\cr & \Sigma_R = \sqrt{w} \left(\matrix{
~A_R J_{\nu-1/2}(w) - B_R J_{-\nu + 1/2}(w)~ \cr ~A_R
J_{\nu+1/2}(w) + B_R J_{-\nu - 1/2}(w)~ \cr} \right) ~~~~~~~~ z >
0 }}
where $w = m (|z| + L)$ and $\nu = M_F L$. Similar fermion
spectroscopy in warped geometries has been analyzed in
\refs{\ygmn,\pomarol}.

We take $l_1 \ge l_2$ for ease of computation \dkklsII, and
set $q_1 = 0$. The
boundary conditions on the branes remove the fermion zero mode
from the spectrum. At low energies, the states \bulkfersol\ break
up into left-localized states, for which $A_R, B_R, B_L \sim
(mL)^{2\nu} A_L$, and right-localized states, for which $A_L,B_L,
B_R \sim (mL)^{2\nu} A_R$. The masses and mass splittings of the
left-localized states are: \eqn\susmasfer{m_L \sim [{(\nu -1)\pi
\over 2} + n\pi] {1 \over l_1} ~~~~~~~~~~~~ \delta m_L = \pm {\cal
O}(1) {q_2 \over l_1}\ ,} while for the right-localized states:
\eqn\susmasferr{m_R \sim [{(\nu -1)\pi \over 2} + n\pi] {1 \over
l_2} ~~~~~~~~~~~~ \delta m_R = \pm {\cal O}(1)
      {q_2 \over l_2} (mL)^{4\nu}\ . }

The mass splittings on the right IR brane arise from loops of the
bulk modes which couple to the SSM. The couplings in the 4d
effective action are determined by their coupling in five
dimensions, given by the appropriate (fractional) power of
$1/M_5$, and by giving canonical normalization to the kinetic term
of the 4d fields.  The latter requires rescaling by powers of the
warp factor $a$, and by the wavefunction of bulk modes evaluated
at the IR brane. Thus the Yukawa couplings of left and right
localized bulk fermions to fields on the right IR brane are
\eqn\fercoups{ g^L{}_{R~4D} = {{\cal O}(1)\over \sqrt{M_5L}}
\sqrt{l_2\over l_1} (mL)^{2\nu} ~~~~~~~~~~~~~~ g^R{}_{R~4D} =
{{\cal O}(1)\over \sqrt{M_5L}} \sqrt{l_2 \over l_2 +
(mL)^{4\nu}l_1}} respectively.  The coupling of left-localized
modes to the right IR brane is suppressed by a barrier penetration
factor $(mL)^{2\nu}$ relative to that of the right-localized
modes.

The effective SUSY breaking scale in the SSM is set by the gaugino
mass. To generate such a mass, we must include a massive $M\sim
1/L$ adjoint scalar on the right IR brane, coupling to the gaugino
and the bulk fermion (as the bulk modes are gauge singlets). We
will choose $M\sim M_{GUT}$ to preserve gauge coupling
unification. Summing over the one-loop contributions from the bulk
KK modes, and {\it then} performing the four-dimensional loop
momentum integrals \mirpe, gives a gaugino mass of \dkklsII:
\eqn\gaumapr{ m_{g} \sim {\cal O}(1)  \Bigl({ q_2 \over  M_5L
}\Bigr)^2 {1\over M l_1} {1\over l_2 + (L/l_2)^{4\nu} l_1} \Bigl(
{ L \over l_2} \Bigr)^{4\nu} \, .}

The SSM cutoff is set by the conformal distance of the SSM to the
UV brane $l_2^{-1}~\sim~10^{15}~GeV$.  Squark masses are generated
by radiative corrections including gaugino loops. They start out
close to zero in the UV, and rise via the RG flow in the IR. As a
result, $m_{sq} \sim TeV$, and is comparable to the gaugino mass,
as in no-scale models \noscale.
Our model predicts a gravitino with mass: \eqn\gravmass{ m_{3/2} =
{\cal O}(1) {q_2 \over M_4^2 l_1^3}\, .} This mode is lighter than
other gravitino KK modes, whose masses are $\sim l_1^{-1},
l_2^{-1}$, because it is protected by SUSY.

Tunnelling suppression  produces large mass hierarchies without
much effort. For example, take $M_5 \sim 10^{16} GeV$, $L \sim
5/M_5$, $l_2 \sim 5L$, $M\sim 1/L$ and $q_2 \le {\cal O}(1)$. The
tunnelling suppression coefficient $\nu$ and the SUSY breaking
scale $l_1^{-1}$ must be chosen so that $m_g \sim TeV$. For
$\nu=1$, i.e. with little tunnelling suppression, the required
SUSY breaking scale is low, $l_1^{-1} \sim 10^{10} GeV$. This
scale implies a micron-range gravitino mass $m_{3/2} \sim eV$. If
$\nu=3$, the SUSY breaking scale should be $l^{-1}_1 \sim 3\times
10^{13} GeV$, closer to the unification scale. The induced
gravitino mass is $m_{3/2} \sim 270 GeV$.

There are also model-independent gravity \refs{\hs,\suGUT} and anomaly \rszero\
mediated contributions to the sparticle masses that are bounded by $m_{3/2}$.
They are subdominant to the tunneling mediated contributions
as long as $l_1 > l_2 \bigl({l_2 \over L}
\bigr)^{2\nu + 1/2} {1 \over M_4l_2}$ \dkklsII.

The hierarchies we produce do not originate from the AdS scaling
as in \RSI. In our case the cutoff on the SSM brane is $M_{GUT}$.
Furthermore, our effect would persist with slight modifications
given any warp factor which raises a barrier between different
throats.

Finally, tunnelling suppression may be used to explain other small
numbers such as neutrino masses and super-weakly coupled particles
\dkklsII.

\newsec{Conclusion}

In this note we have studied string phenomenology in a new
calculable regime which can arise when there are many branes
transverse to the compactification manifold. Such
compactifications generate brane throats which provide a new
mechanism for producing small numbers in Nature, by
utilizing the tunnelling suppressed couplings of IR sectors
separated by a potential barrier.


~
\medskip
\centerline{\bf{Acknowledgements}} It is a pleasure to thank N.
Arkani-Hamed, S. Giddings, J. Maldacena, A. Peet, M. Peskin, J.
Polchinski, A. Pomarol, M. Schmaltz, S. Shenker, P. Steinhardt, A. Strominger,
L. Susskind, S. Thomas, H. Tye and H. Verlinde for interesting discussions.
This work was supported in part by the NSF grants PHY-99-07949 and
PHY-9870115 and the DOE under contract DE-AC03-76SF00515. We thank
the ITP at Santa Barbara for hospitality.  A.L. thanks the CGTP at
Duke University for hospitality and support from NSF grant
DMS-0074072. S.K. was supported in part by a Packard Fellowship
and a Sloan Fellowship, and E.S. was supported in part by a Sloan
Fellowship and a DOE OJI grant.

\listrefs
\end